\documentclass[sigconf, nonacm]{acmart}
\usepackage{algorithm}
\usepackage{algpseudocode}
\usepackage{amsmath}
\usepackage{caption}
\usepackage{graphicx}
\usepackage{subcaption}
\usepackage{balance}

\usepackage{tabularx}
\usepackage{multirow}

\usepackage{array, rotating}
\usepackage{multirow, makecell}

\newcommand{\probP}{\text{I\kern-0.15em P}}
\algrenewcommand\algorithmicrequire{\textbf{Input:}}
\algrenewcommand\algorithmicensure{\textbf{Output:}}

\usepackage{xcolor}
\newcommand{\hblue}[1]{\textcolor{black}{#1}}

\newcommand\vldbdoi{10.14778/3659437.3659458}
\newcommand\vldbpages{2064 - 2076}
\newcommand\vldbvolume{17}
\newcommand\vldbissue{8}
\newcommand\vldbyear{2024}
\newcommand\vldbauthors{\authors}
\newcommand\vldbtitle{\shorttitle} 
\newcommand\vldbavailabilityurl{https://github.com/fzirak/SeLeP.git}
\newcommand\vldbpagestyle{empty} 

\newcommand{\far}{\textsf{SeLeP}}

\newcommand{\centered}[1]{\begin{tabular}{c} #1 \end{tabular}}

\begin{document}
\title{{\far}: Learning Based Semantic Prefetching for Exploratory Database Workloads}

\author{Farzaneh Zirak}
\affiliation{%
  \institution{University of Melbourne}
  \city{Melbourne}
  \country{Australia}
}
\email{fzirak@student.unimelb.edu.au}

\author{Farhana Choudhury}
\orcid{0000-0001-6529-4220}
\affiliation{%
  \institution{University of Melbourne}
  \city{Melbourne}
  \country{Australia}
}
\email{farhana.choudhury@unimelb.edu.au}

\author{Renata Borovica-Gajic}
\orcid{0000-0003-3503-4123}
\affiliation{%
  \institution{University of Melbourne}
  \city{Melbourne}
  \country{Australia}
}
\email{renata.borovica@unimelb.edu.au}

\begin{abstract}
    Prefetching is a crucial technique employed in traditional databases to enhance interactivity, particularly in the context of \textit{data exploration}. Data exploration is a query processing paradigm in which users search for insights buried in the data, often not knowing what exactly they are looking for. Data exploratory tools deal with multiple challenges such as the need for interactivity with no a priori knowledge being present to help with the system tuning. The state-of-the-art prefetchers are specifically designed for navigational workloads only, where the number of possible actions is limited. The prefetchers that work with SQL-based workloads, on the other hand, mainly rely on data logical addresses rather than the data semantics. They fail to predict complex access patterns in cases where the database size is substantial, resulting in an extensive address space, or when there is frequent co-accessing of data. In this paper, we propose {\far}, a semantic prefetcher that makes prefetching decisions for both types of workloads, based on the encoding of the data values contained inside the accessed blocks. Following the popular path of using machine learning approaches to automatically learn the hidden patterns, we formulate the prefetching task as a time-series forecasting problem and use an encoder-decoder LSTM architecture to learn the data access pattern. Our extensive experiments, across real-life exploratory workloads, demonstrate that {\far} improves the hit ratio up to 40\% and reduces I/O time up to 45\% compared to the state-of-the-art, attaining 96\% hit ratio and 84\% I/O reduction on average. 
\end{abstract}

\maketitle

\pagestyle{\vldbpagestyle}
\begingroup\small\noindent\raggedright\textbf{PVLDB Reference Format:}\\
\vldbauthors. \vldbtitle. PVLDB, \vldbvolume(\vldbissue): \vldbpages, \vldbyear.\\
\href{https://doi.org/\vldbdoi}{doi:\vldbdoi}
\endgroup
\begingroup
\renewcommand\thefootnote{}\footnote{\noindent
This work is licensed under the Creative Commons BY-NC-ND 4.0 International License. Visit \url{https://creativecommons.org/licenses/by-nc-nd/4.0/} to view a copy of this license. For any use beyond those covered by this license, obtain permission by emailing \href{mailto:info@vldb.org}{info@vldb.org}. Copyright is held by the owner/author(s). Publication rights licensed to the VLDB Endowment. \\
\raggedright Proceedings of the VLDB Endowment, Vol. \vldbvolume, No. \vldbissue\ %
ISSN 2150-8097. \\
\href{https://doi.org/\vldbdoi}{doi:\vldbdoi} \\
}\addtocounter{footnote}{-1}\endgroup

\ifdefempty{\vldbavailabilityurl}{}{
\vspace{.3cm}
\begingroup\small\noindent\raggedright\textbf{PVLDB Artifact Availability:}\\
The source code, data, and/or other artifacts have been made available at \url{\vldbavailabilityurl}.
\endgroup
}

\section{Introduction}
    Exploring massive amounts of data to extract (unknown) information is a query processing paradigm called \emph{data exploration}~\cite{StratosExploration, noDBSIGMOD2012, idreos2015overview}. The growth in data collection ability in recent decades has led to providing larger and more detailed datasets in both sciences and businesses~\cite{ScientificDataManagementGray}. Consequently, the popularity of data exploration has significantly increased, giving rise to the need for database systems tailored to its specific requirements, such as responding interactively and adapting to the shifts in the users' workload~\cite{ScientificDataManagementGray, ResearcherGuide}.

    It has been shown that during exploratory browsing, the interaction response times should be bounded within 500 ms, since additional delays drastically reduce the rate by which users make observations, draw generalizations, and generate hypotheses \cite{liu2014latencyeffects}. %
    However, given the exponential growth in the amounts of generated data, responding to queries over such large data sets with a subsecond latency has become a tall order for the traditional database management systems (DBMS)~\cite{ResearcherGuide}.
    As a result, fetching the data \emph{prior} to the user request is one of the approaches added to the traditional DBMSs to address the issue of non-interactive performance over ever-growing data sets. Usually, an exploratory session contains multiple queries often with pauses between them as users contemplate the next query, since the result of each query affects the formulation of the next one~\cite{doshi2003prefetching, StratosExploration, ResearcherGuide}. After each query, while the user is interpreting the recently obtained data and preparing the subsequent request, the unoccupied system can try to predict the data likely to be requested next and fetch it into the cache (if not already there) before the user asks for it. Thus, when the data is requested later, the user perceives a lower response time. 
    
    In exploratory tools, prefetching has mainly been used in visual exploration in which the operations are limited to zooming and panning \cite{doshi2003prefetching, atlas, tauheed2012scout, battle2016prefetching, wan2018learning}. ForeCache \cite{battle2016prefetching} and Scout \cite{tauheed2012scout} utilize data characteristics in their prediction algorithm in addition to the spatio-temporal locality of the data and prove that data-driven models can achieve a better access prediction accuracy and cache hit ratio compared to the other action-based prefetchers. However, none of these techniques can work with query workloads comprising full-blown SQL queries since they only examine and pick data in the vicinity of the current accessed area to prefetch.  

    Recently proposed memory prefetchers \cite{bera2021pythia, 2020delta_lstm, 2023sgdp} have demonstrated that machine learning techniques have great power in automatically learning patterns from a sequence of data addresses. While these prefetchers can be applied to databases, they often overlook the co-accessing of data. In other words, when an application requests $n$ data pages simultaneously, they treat it as a sequence of length $n$ rather than a group. Consequently, prefetching decisions depend on the order of the generated sequence, and the system does not consider the dependencies between all accessed addresses. Furthermore, in our experiments we observe that these models mainly focus on predicting a single data access and do not efficiently handle multiple prefetch decisions in an interactive pace. Additionally, most of these prefetchers lack the utilization of data semantics and instead rely on logical block addresses (LBA) or other environmental parameters to learn patterns and make predictions.

    Recurrent neural networks (RNNs), and more specifically, Long Short-Term Memory (LSTM) models, are the most popular learning methods used in memory prefetchers~\cite{bera2021pythia, 2020delta_lstm, 2023sgdp}. Taking inspiration from the learning-based and the data-driven prefetchers, we propose {\far} -- a semantic-based prefetcher designed for both visual and SQL-driven exploratory workloads. In our approach, we treat the prefetching problem as a time series forecasting problem, where each element in the time series represents a group access to the data. We employ an encoder-decoder LSTM architecture to tackle this problem. However, unlike past efforts that use logical addresses of the accessed data~\cite{2020delta_lstm, 2023sgdp}, {\far} learns and predicts the sequence of \emph{semantic-based encodings} of the data values.
    
    {\far} obtains a vector representation for each physical block by employing deep learning feature extraction methods on its data values. These vector values capture data semantics and serve as the key components of {\far}, analogous to logical block addresses in memory prefetchers. As demonstrated in \cite{battle2016prefetching} and \cite{tauheed2012scout}, considering data semantics rather than solely addresses can be beneficial since blocks are typically accessed based on the values they contain. For instance, when querying a dataset, the retrieved blocks contain certain values, leading to their co-accessing. 
     
    Due to the challenges involved with access prediction among a massive number of blocks, {\far} groups the blocks that are frequently accessed together into partitions and prefetches data in units of partitions instead of blocks. Additionally, by using these partitions, {\far} can more effectively prefetch data that are related to each other and are likely to be accessed together, which can enhance the prefetching strategy and improve overall performance. 
     
    For each partition, {\far} creates a matrix representation using the vector encodings of their assigned blocks. It also uses the matrix encodings to represent a query and employs an LSTM model on the sequence of query encodings to predict the subsequent partition accesses. With the query encodings, {\far} can consider a group of accessed blocks and does not need to define an ordered sequence on the blocks accessed by a single query.
       
    The output of the LSTM is the probability of each partition being accessed subsequently. Using this prediction model, {\far} selects and brings the partitions most likely to be accessed into the cache.
    The contribution of this paper are summarized as follows:
    \vspace{-0.2em}
    \begin{itemize}
        \item To the best of our knowledge, we propose the first prefetcher that leverages data semantics by encoding data blocks and makes predictions based solely on the encoding sequences.
        \item We propose a prefetcher that unlike state-of-the-art can cater for both visual and SQL-based exploratory workloads.   
        \item We formalize the prefetching problem as a times series forecasting problem and demonstrate that the prediction models with semantics inputs can outperform the models that utilize logical data addresses.
        \item We conduct an extensive experimental evaluation across both types of workloads gathered from real-world scientific datasets, as well as a publicly available industrial benchmark, and show that {\far} outperforms the state-of-the-art prefetchers by up to 40\% in hit ratio, achieving 95\% hit ratio on average across all experiments.
        
    \end{itemize}

    \begin{figure*}
        \centerline{\includegraphics[width=0.98\textwidth, height=5.8cm]{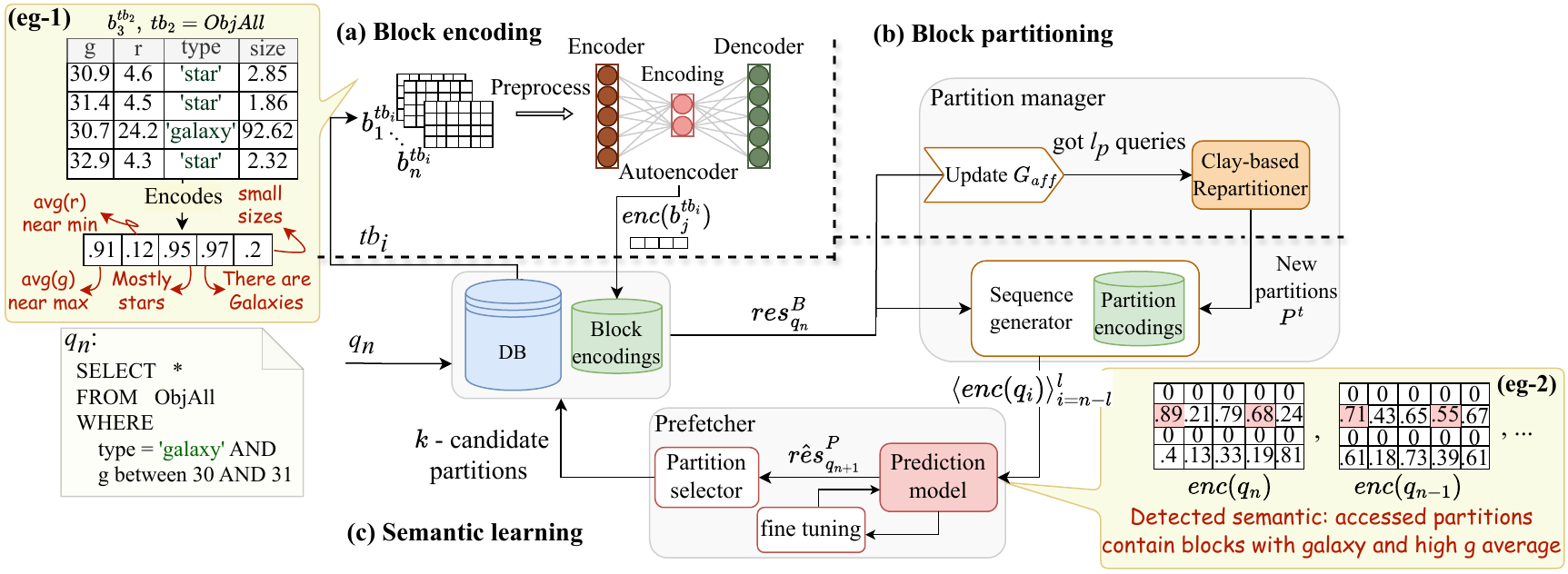}}
        \vspace{-0.5em}
        \captionsetup{belowskip=-12pt}
        \caption{\hblue{System architecture of {\far} representing: (a) the block encoding component generating block vector encodings using autoencoders, (eg-1) a simplified example for block encoding, (b) the block partitioning component clustering the blocks, (c) the semantic learning component predicting the partition access pattern based on semantic relationship of query encodings, as illustrated in (eg-2). After initial training, (b) and (c) continuously perform prefetching.}}
        \label{fig_model_overall}
    \end{figure*}

\section{Problem Formulation}\label{sec:problem_formulation}
    The core of the semantic prefetching problem is building a prediction model that captures the inter-dependencies of the previously requested data and anticipates future accessing data to fetch it into the cache prior to the request of an upcoming query. We now formulate semantic prefetching as a time-series forecasting problem.

     Let set $B$ represents all blocks (i.e., pages of the data) present in the database, $b^{tb_i}_n \in B$ symbolizes n\textsuperscript{th} block of table $tb_i$ in the database. Consider a sequence of queries $\langle q_i\rangle_{i=1}^{n}=\langle q_1, q_2,\dots,q_n\rangle$ executed within a session, with each query execution time considered as a distinct timestep. The set $res^B_{q_i} \subseteq B$ represents the blocks accessed by query $q_i$. 
     
     Given the sequence of $\langle res^B_{q_i}\rangle_{i=1}^{n}$, the non-semantic prefetching problem involves predicting and fetching the $res^B_{q_{n+1}}$ regardless of the data values and considering the LBAs and other environmental parameters. However, our objective is to determine the inter-dependencies among query result sets of the $n$ most recent executed queries, and anticipate the data that will be accessed next. Therefore, we define the semantic prefetching as follows:    
    \begin{definition}[Semantic Prefetching]
        \textit{Having a sequence of query result blocks corresponding to the $n$ most recently executed queries $\langle res^B_{q_i}\rangle_{i=1}^n$, capture the connection between their data values to predict and fetch the upcoming block access request $res^B_{q_{n+1}}$.}
    \end{definition}
  
\section{{\protect\far}: Learning-based Semantic Prefetching framework}  
    In this section, we provide an overview of the proposed learning-based semantic prefetcher, followed by an in-depth explanation of its key components. We start by presenting an overview in \S \ref{sec:arch_overview}. Then, we explore the details of the block encoding process in \S \ref{sec:data_encoding}, the partitioning process in \S \ref{sec:data_partitioning}, and the semantic learning and prediction model in \S \ref{sec:semantic_learning}, making complete picture of {\far} operation.

 \subsection{{\far} Overview}\label{sec:arch_overview}
    {\far} is a learning-based prediction framework that extracts the semantic relationships within the previously requested data and outputs the future data accesses. Due to the small size of blocks (8kB to 32kB) and extensive database sizes, a substantial volume of blocks is present in the system. The numerous number of blocks leads to a considerable number of potential outcomes for the prediction model, which can reduce its accuracy. Hence, we adopt a dynamic clustering approach to group related blocks into more sizable sets, called partitions, and use them instead of blocks in the prefetching process. The impact of partitioning approach is assessed in \S\ref{sec:sensitivity_test}.
    
    
    Let $P^t$ denote the set of partitions generated on the blocks at timestep $t$, where $p^t_i$ and $bp_i^t = \{b|b\in B \wedge b \in p^t_i\}$ represent an individual partition and its set of assigned blocks respectively. To simplify the presentation and since the partitions do not change frequently, we omit the timestep $t$ indices in the reminder of this paper\footnote{We however do test the impact of partition changes in \S \ref{sec:adaptivity_test}.}. Using the partitions, we define the set $res^P_{q_i} \subseteq P$ as the partitions accessed by query $q_i$. 
    
    To grasp the data semantics, we employ feature extraction methods to encode the data, and express each query by aggregating the encoding of its result set, represented as $enc(q_{i})$. By harnessing these query encodings, the partition level semantic prefetching problem transfers to a time series forecasting problem with query encoding sequence $\langle enc(q_{i})\rangle_{i=1}^n$ as input and $res^P_{q_{n+1}}$ as the output.
    
    The overall framework of our system is depicted in Figure \ref{fig_model_overall}. It comprises three distinct components: block encoding as a preprocessing step, and continuous block partitioning and semantic learning. Before the system can utilize these components, a data preparation step is required. During the data preparation step, we execute and collect result of a set of queries $q$ selected for training. These result blocks are used to create the training workloads for warming up the system and training the model.
    
    Within the block encoding module, Figure \ref{fig_model_overall}(a), an Autoencoder \cite{rumelhart1985learning} is employed on each table to generate vector representations for its blocks. Prior to utilizing the Autoencoder for block encodings, the data is preprocessed based on properties of its corresponding table such as the number of columns and range of values (with more details provided in \S \ref{sec:data_encoding}). This process is executed once, and the block encodings are stored for future reuse. \hblue{Figure \ref{fig_model_overall}(eg-1) depicts a simplified block encoding example, showing that the encoded vector provides a concise representation of the block's characteristics.}
    
    The block partitioning component, depicted in Figure \ref{fig_model_overall}(b), groups blocks frequently co-accessed by the workload into the same partition. In other words, this component clusters the blocks, and those co-accessed by many queries are more likely to be assigned to the same partition. We extend the Clay partitioning algorithm \cite{2016clay} (details provided in \S \ref{sec:data_partitioning}), which employs an affinity graph to represent the co-access frequency between pairs of blocks. During the initial phase, we apply this algorithm on the queries selected for training the prediction model to create and warm up the partitions. Later, after the complete system is prepared for prefetching, we leverage this component to update the affinity graph and the partitions. We undertake iterative repartitioning following the reception of a certain number of queries to ensure accurate capture of the co-access frequency inherent in the most up-to-date workload.

    Once the blocks are encoded, and the partitions are established, we have the required data to train our prediction model in the semantic learning component (Figure \ref{fig_model_overall}(c)). To prepare the train and test data, we use the $res^B_{q}$ generated in data preparation step and $P$ formed in block partitioning to obtain $res^P_{q}$ and calculate $enc(q)$ for $q$ in the training workloads. We consider each $l$ (named lookback) consecutive query and generate $\langle enc(q_{i})\rangle_{i=n-l}^n$ as the model input. For the output, we employ a bitmap string where each bit indicates the presence or absence of a partition in $res^P_{q_{n+1}}$.

    After the model is trained, we can use the system to make prefetch decision. In this stage, when a new query is executed, the system determines its list of accessed partitions, updates the affinity graph, and generates $\langle enc(q_{i})\rangle_{i=n-l}^n$ using the previous queries to input the model and extract the semantic relation between the queries. \hblue{Figure \ref{fig_model_overall}(eg-2) illustrates a simplified example of semantic relationship between requests.} The model output can be interpreted as the probability of accessing each partition in the subsequent request. Consequently, the system identifies the top-$k$ values, which correspond to the partitions most likely to be requested next, and retrieves the blocks associated with these partitions into the cache.

    The last two components are employed continuously: the block partitioning component updates the affinity graph based on the received queries. Once a certain number of queries ($l_p$) are received, i.e., the repartitioning threshold is reached, the partitioning algorithm is applied to the latest version of the affinity graph. During repartitioning, blocks are shuffled between partitions, and the partition encodings are modified. Hence, the accuracy of the prediction model may reduce after repartitioning. 
    To address this, after each repartitioning, we proceed to fine-tune the prediction model. This process involves using the new partition encodings and the set of queries executed since the last repartitioning. 
    
    Table \ref{tab:notations} lists the frequently used notations in this paper.
    \begin{table}[htbp]
    \caption{\hblue{Frequently Used Notations}}
    \vspace{-0.8em}
    \begin{center}
            \scalebox{0.9}{
    \begin{tabular}{|c|c|}
        \hline
        \textbf{Symbol}&\textbf{Definition}\\
    	\hline
        $b^{tb_i}_n$ & $n^{th}$ block in table $tb_i$ \\
        \hline
        $bp_i$ & Blocks allocated to partition $p_i$ \\
        \hline
        $res^B_q$, $res^P_q$ &\centered{Set of blocks and  partitions accessed by query $q$} \\
        \hline
        $enc(b)$, $enc(p)$ & Block vector encoding, partition matrix encoding
        \\
        \hline
        $l_{be}$ & Length of block encodings \\
        \hline
        $l$ &\centered{Lookback. Sequence length considered by the\\model for prediction} \\
        \hline
        $l_p$ &\centered{Repartitioning threshold
        } \\
        \hline
        $k$ & Prefetch size in unit of partitions \\
        \hline 
    \end{tabular}
     }
     \label{tab:notations}\vspace{-1em}
    \end{center}
      \vspace{-0.6em}
    \end{table}

\vspace{-0.05em}
 \subsection{Block Encoding}\label{sec:data_encoding}   
    The purpose of semantic prefetching is to incorporate the actual data values, rather than just block addresses, in access pattern recognition. Since each data block can contain hundreds of values, it is challenging to identify relationships among such vast amounts of data in a series of blocks. Therefore, we need to have a concise representation for each block which captures the essential and distinctive characteristics of its data. To this end, we consider each block of data as a big matrix and encode each matrix to a vector with a block encoding length, denoted as $enc(b_n^{tb_i})$ and $l_{be}$ respectively.

    As we lack prior knowledge about the most crucial parts of the data, we address the block encoding problem as an unsupervised feature extraction task, employing an autoencoder model to extract information from the data. Autoencoders \cite{rumelhart1985learning} are unsupervised neural network architectures specifically designed for feature extraction and representation learning. These models, used for compressing and encoding various data types including images \cite{2020imgAutoencoder}, excel at handling large matrices, making them well-suited for our encoding problem. During the block encoding process, we start by preprocessing the data of each table, followed by training an autoencoder to learn the encoding of its blocks. 
    
    \subsubsection{Data Preprocessing}
    A key challenge in encoding blocks into fixed-size vectors when using autoencoders is to convert all data types into numeric values usable by the model. To address this, we convert date and time values to Unix timestamp and other non-numeric types to text, which later are encoded to numbers using Word2vec models~\cite{word2vec}. We opt for training a Word2vec model for each table, since not all texts, including those originally in character type, are meaningful. For instance, `B\&D\_NaD' is a star name in the Sloan Digital Sky Survey (SDSS) \cite{abazajian2009sdss} DR7 dataset. Hence, each string is treated as a sentence and fed into the Word2vec model, resulting in each column being converted into 8 new numeric columns.

    Next challenge is encoding tables with a wide structure (i.e. with hundreds of columns), while typically many columns carry insignificant or repetitive values. For instance, within the SDSS dataset, several tables exceed 500 columns, some filled with default or null values, or displaying variations of one attribute observed at different times. The presence of many such columns can increase the loss value of the encoder and result in a less representative encoding. In addition, columns with vast range of numerical values can hinder the encoding process by causing scaling challenges in the activation functions and loss computation parts \cite{huang2023normalization}.

  \begin{table}[htbp]
    \caption{Average block encoding time, hit ratio at $k=40$ and ROC AUC score of the system with different block encoders on the SDSS dataset. The best and second-best performance are highlighted in bold and underlined, respectively.}
      \vspace{-0.5em}
    \begin{center}
            \scalebox{0.84}{
  
    \begin{tabular}{|*6{c|}c|}
    	\hline
    	\multicolumn{2}{|c|}{Test}&SLP&CNN&MLP\scriptsize{(\far)}&LSH&AggPCA\\
    	\hline	\multirow{2}{*}{{s-reg}}
    	&HR&93.97&\underline{94.52}&\textbf{95.19}&85.90&90.44\\
    	 &\small{ROC AUC}&0.854&\underline{0.862}&\textbf{0.891}&0.835&0.855\\
    	\hline	\multirow{2}{*}{{s-rand}}
    	&HR&\underline{96.98}&96.51&\textbf{97.94}&91.06&96.67\\
    	 &\small{ROC AUC}&0.851&0.840&\textbf{0.881}&0.841&\underline{0.855}\\
    	\hline	\multirow{2}{*}{{m-reg}}
    	&HR&\textbf{99.90}&\textbf{99.90}&99.81&\textbf{99.90}&\underline{99.86}\\
    	 &\small{ROC AUC}&0.988&\textbf{0.992}&\underline{0.991}&0.990&\underline{0.991}\\
    	\hline	\multirow{2}{*}{{m-rand}}
    	&HR&\underline{99.76}&\textbf{99.88}&\underline{99.76}&\underline{99.76}&\textbf{99.88}\\
    	 &\small{ROC AUC}&0.985&0.986&\underline{0.987}&\textbf{0.989}&0.986\\
    	\hline	\multirow{2}{*}{{mj-reg}}
    	&HR&\underline{96.20}&96.02&\textbf{96.98}&92.07&89.12\\
    	 &\small{ROC AUC}&0.834&\underline{0.839}&\textbf{0.884}&0.782&0.829\\
    	\hline	\multirow{2}{*}{{mj-rand}}
    	&HR&93.09&\underline{95.00}&\textbf{95.67}&91.09&89.57\\
    	 &\small{ROC AUC}&0.859&\underline{0.882}&\textbf{0.893}&0.821&0.840\\
    	\hline	\multirow{2}{*}{{f-SDSS}}
    	&HR&78.55&\underline{85.95}&\textbf{86.17}&74.55&81.19\\
    	 &\small{ROC AUC}&0.680&\underline{0.759}&\textbf{0.785}&0.655&0.732\\
    	\hline
    	\multicolumn{2}{|c|}{Encoding Time}&\underline{1868.35}&4357.3 & 2898.17&2983.12&\textbf{570.51}\\
    	\hline
    \end{tabular}
     }
     \label{tab: encoding_justification}
     \vspace{-1.3em}
    \end{center}
    \end{table}

    To address these issues, we apply min-max normalization (Equation (\ref{min-max_eq})) on each column individually, ensuring that all values are within the range of -1 to 1. Subsequently, we employ principal component analysis (PCA) \cite{pearson1901liii} to reduce table sizes and filter their columns before feeding them into the autoencoder. PCA identifies a meaningful basis to transform the dataset into a lower dimension by computing new orthogonal axes called principal components. With these axes, we can map the original data to a lower dimension while preserving its underlying pattern.
    \begin{equation}
         x_{normalized} = \frac{x-min(X)}{max(X) - min(X)}\times 2-1\label{min-max_eq}
    \end{equation}
    

    \subsubsection{Autoencoders}
    Autoencoders consist of two main components: the encoder that encodes the data into a compressed size which is called latent space or the encoding, and the decoder that takes the produced encodings and aims to regenerate the original data. By using autoencoders for block encoding two questions arise:\textit{ (i) How well does autoencoder perform compared to heuristic-based dimensionality reduction approaches?} and \textit{(ii) Which layer architecture performs better in the autoencoder?}
    
    To answer these questions, we conduct experiments with various prefetch sizes ($k$) using several block encoders on the SDSS dataset (workloads are explained in \S \ref{sec:workloads}). We design three variants of autoencoders: one with a single dense (a.k.a fully connected) layer (SLP), another with two dense layers (MLP), and one with two convolutional neural network layers (CNN) \cite{cnn}, and two heuristic-based alternatives: one utilizing PCA for dimension reduction and row-wise feature aggregation (AggPCA), and the other employing Local Sensitivity Hashing (LSH) \cite{lsh} for data compression.
    
    Table \ref{tab: encoding_justification} presents the hit ratio (HR) for each system at $k=40$. Additionally, the Receiver Operating Characteristic Area Under the Curve (ROC AUC) score is provided to demonstrate each encoder's overall performance across all $k$ values. A higher score indicates a system's capability to achieve a greater hit ratio within the tested $k$ range. We observed that the performance of autoencoders is heavily influenced by the characteristics of the block data.
    
    While heuristic-based methods can effectively compress data with simple patterns, they often fail to extract their semantics which negatively affects the performance in more complex workloads (mj-rand, mj-reg, and f-SDSS). Similarly, SLP struggles with high-dimensional or complex data, but excels at encoding tables with few columns, achieving high performance on workloads accessing these tables (e.g. s-rand). Although CNN and MLP can both handle complex patterns and achieve a high hit ratio, we choose MLP since it offers a simpler model and faster encoding (presented in Table \ref{tab: encoding_justification}).


    
    
    As depicted in Figure \ref{fig_model_overall}(a), an autoencoder is created for each table $tb_i$, trained with the table's blocks. Subsequently, the trained encoder converts blocks into vectors, represented as $enc(b^{tb_i}_n)$, which will be stored in the system for later use.

 \subsection{Block Partitioning}\label{sec:data_partitioning}   
    As mentioned earlier, due to the wide table schema and large numerical values, a single data record could be as large as a few kilobytes. Therefore, prefetching a single 8kB block of data results in caching very few data records. Additionally, the large size of data objects and the extensive size of the database result in a tremendous number of blocks present in the system. This considerable size causes challenges in accurately predicting the subsequent data accesses. 
    
    To address these issues, we cluster blocks into partitions, which are more sizeable sets (in particular 128 blocks in our experiments), and use the partition number to refer to a data access. Figure \ref{fig:sensitivity}(d) depicts the effect of partition size on prefetching performance. We form the initial partitions by grouping the consecutive blocks of the tables. Subsequently, the system periodically performs repartitioning to cluster the blocks that are commonly accessed together in queries. The repartitioning process is done based on the most recently received workload represented by a graph data structure.

    \subsubsection{Construction of Affinity Graph}
    To group the frequently co-accessed blocks, we construct a block access graph called affinity graph $G_{aff}$, with nodes representing the blocks, edges illustrating co-accessing relation and the weight of the edge reflecting the frequency of the corresponding nodes being co-accessed. To create this graph, a node is added for a block only if it is accessed at least once. In data exploration, despite the substantial dataset sizes, only small portions of the data are of interest to the users~\cite{ResearcherGuide, StratosExploration, idreos2015overview, noDBSIGMOD2012}. In other words, there may be many blocks that never get to be requested or accessed. Thus, though the datasets contain a large number of blocks, the size of the graph can remain relatively small.
    
\begin{figure}
       \centerline{\includegraphics[width=0.84\linewidth]{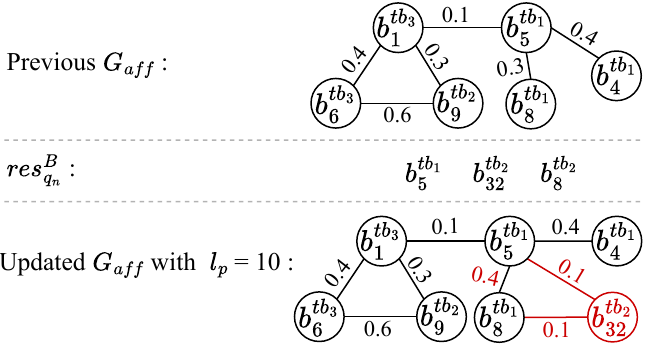}}
         \vspace{-0.5em}
        \captionsetup{belowskip=-10pt}
        \caption{An example of affinity graph modification upon receiving a new query which accesses block $b_{32}^{tb_2}$ for the first time ($l_p$ is set to a very small value to simplify the example)}
        \label{affinity_graph}
         \vspace{-0.2em}
    \end{figure}

    The system observes the queries and modifies $G_{aff}$ for a batch of queries with length $l_p$. For each query $q$ in the batch, it finds $res^B_{q}$ and passes this list to the partition manager. For every accessed block, the system creates a node in the graph (if it does not already exist) and includes an edge connecting the nodes corresponding to each pair of blocks (if the edge does not already exist). The weight of the edges will increase by $1/l_p$ to demonstrate how frequently a pair has been co-accessed during the current batch. Figure \ref{affinity_graph} provides an example of the affinity graph modification process with a very small value for $l_p = 10$. Once a complete batch of $l_p$ queries is received, the system triggers a repartitioning process to update the partitions based on the most recent co-accesses represented by $G_{aff}$. 
    
    \subsubsection{Clay-based Partitioning}
    Using $G_{aff}$, the partitioning problem can be transformed into a graph partitioning problem. Among the methods proposed for graph partitioning, Clay~\cite{2016clay} stands out for its application in load balancing distributed databases. It monitors the data accesses and generates a heat map similar to our affinity graph. Its repartitioning module is triggered when a server's load reaches the maximum load threshold $\theta$. At the time of repartitioning, Clay selects a group of data records causing the overload containing the most accessed record and its relative neighbors. It then determines the best server to migrate the chosen records, ensuring the destination server isn't overloaded. If no suitable destination is found, it adds a new empty server to the system.

    We have adopted the Clay technique since it focuses on minimizing distributed transactions between partitions, which is different from most graph partitioning techniques that prioritize balancing partition accesses. Also, it is an online partitioning approach that migrates data in dynamically generated groups rather than static sets of blocks. To tailor Clay to our problem, we have performed the following modifications:
    \begin{itemize}
        \item Clay considers the rate of access to individual blocks as well as their co-accessing rate to calculate the load of each partition. However, in our case, we only focus on the co-accessing load and disregard the number of times each individual block has been accessed. Therefore, the load for each partition can be calculated by:
        \begin{equation}
            load_{p_i} = \sum_{v \in p_i\ \wedge\ u\notin p_i} w(edge(v,u))\ .\ k_w \label{eq_load}
        \end{equation}
        where $u$ and $v$ are two nodes in $G_{aff}$ linked by an edge with weight $w$, and $k_w$ is a constant weight indicating how much inter-partition accesses are tolerable in the system.
        
        \item If a proper destination for migrating the overloading nodes cannot be found, Clay adds a new empty server to the system and reruns the algorithm. However, in our case, the number of partitions is fixed, and new partitions cannot be added to the system. To address this restriction, we made multiple modifications. During the initial partitioning, we ensure that no more than 90\% of the partition size is filled. Additionally, after assigning blocks to the partitions, we integrate an additional 5\% of the total number of partitions ($|P|$) as empty partitions into the system\footnote{We measure the impact of each configuration parameter in \S \ref{sec:sensitivity_test}.}. Furthermore, we make $\theta$ dynamic and increase it when we need to add a new partition to the system. It is important to note that maintaining a fixed $\theta$ in the original Clay system can also lead to issues, as a group of data with a size exceeding the maximum server capacity may frequently be accessed.
        
        \item Clay discards the graph after each repartitioning, but we take a different method to retain workload history. Instead of discarding the graph entirely, we scale down the weights by a factor less than 1, preserving historical access information with reduced impact on future partition formations.
    \end{itemize}

    \subsubsection{Partition Encoding}
    Once a new partition set $P$ is formed, the system calculates the partition encodings, labelled as $enc(p)$, by aggregating the encodings of their assigned blocks. However, it is inadvisable to aggregate encoding of blocks that belong to different tables. This is because individual fields within the vector encoding of blocks from different tables have distinct semantic interpretations.
     For instance, in Figure \ref{fig_model_overall}(eg-1), the 5th element of block encodings in $tb_2$ indicates the size of objects, which can be entirely different from the 5th element of encodings in other tables. Hence, we cannot aggregate $enc(b^{tb_i}_n)$ and $enc(b^{tb_j}_m)$ if $i \neq j$.
    
    To mitigate this issue, we encode each partition as a matrix rather than a vector. This matrix is structured with rows corresponding to individual tables within the database and $l_{be}$ columns. During the data preparation step, we enumerate and assign an integer to each table, indicating their corresponding row in matrix encodings. Consequently, within a given partition, the vector encodings of blocks belonging to each table can be aggregated separately and stored in distinct rows of the matrix corresponding to that table. 

    The partition encoding process is outlined in Algorithm \ref{partition_enc_alg}. First, the partition encoding is initiated with an all zero matrix with number of rows equal to the count of tables within the dataset, and $l_{be}$ columns. Subsequently, blocks within each partition are grouped according to their respective tables. For each table $tb_j$ with blocks assigned to the partition, the algorithm calculates an aggregated encoding by averaging the individual block encodings. This aggregated encoding is then stored in the $j^{th}$ row of the matrix.
 
 \vspace{-0.2em}
 \subsection{Semantic Learning}\label{sec:semantic_learning}

    In this section, we describe the LSTM model used for solving the semantic prefetching problem and compare it against some alternative models. LSTM is one of the models within the family of RNNs specifically designed to learn short-term and long-term dependencies among a sequence of data~\cite{hochreiter1997lstm}. Due to its capability to capture temporal patterns, it can be an excellent choice for addressing time-series forecasting problems. \hblue{An RNN-based alternative to LSTM is the Gated Recurrent Unit (GRU) \cite{gru}, which is known for its simplicity and fewer trainable parameters compared to LSTM. The question that arise is: \textit{Which model and what architecture perform better as the prediction model of our semantic prefetcher?}}
    
    We evaluate various prediction models, composed of different learning models, such as a GRU model, an LSTM model, a two-layer LSTM (ML-LSTM) model, a three-layer MLP model, and an encoder-decoder LSTM (ED-LSTM) model, on the SDSS dataset. The results are shown in Table \ref{tab: prediction_justification}. We observe that the MLP model struggles to effectively  capture the query semantics and partition access pattern and performs poorly compared to the other RNN-based models.

    \begin{algorithm}
            \caption{The Partition Encoding Calculation Algorithm}\label{partition_enc_alg}
            \begin{algorithmic}[1]
                \Require Partitions $P$, Block Encodings $\{enc(b)\}$, Count of Tables $n_{tb}$, Block Encoding Length $l_{be}$ 
                \For {$p_i$ in $P$}
                \State $enc(p_i)\leftarrow$ zeros($n_{tb}$, $l_{be}$)
                \State Separate blocks of each table within $bp_i$ 
                \For {$tb_j$ involved in $bp_i$}
                    \State $bp_i^{tb_j}\leftarrow$ blocks of table $tb_j$ in $bp_i$ 
                    \State $aggregated\_enc$ $\leftarrow$ mean($enc(b)$ for $b$ in $bp_i^{tb_j}$)
                    \State  $enc(p_i)[j]$ $\leftarrow$  $aggregated\_enc$
                \EndFor
                \EndFor
            \end{algorithmic}
        \end{algorithm}
     \begin{table}[htbp]
            \vspace{-0.5em}
    
        \caption{Best hit ratio of the system with different prediction models on the SDSS dataset.
            \vspace{-0.8em}
        }
        \begin{center}
                \scalebox{0.92}{
        
             \begin{tabular}{|*6{c|}c|}
            	\hline
            	Test&MLP&GRU&LSTM&ML-LSTM&ED-LSTM\scriptsize{(\far)}\\
            	\hline	s-reg&93.61&93.69&89.32&\underline{94.00}&\textbf{95.19}\\\hline
            	s-rand&95.87&96.83&\underline{97.87}&96.35&\textbf{97.94}\\\hline
            	m-reg&\textbf{99.90}&\textbf{99.90}&\textbf{99.90}&\underline{99.86}&99.81\\\hline
            	m-rand&\underline{99.76}&\textbf{99.88}&99.64&\underline{99.76}&\underline{99.76}\\\hline
            	mj-reg&89.23&91.08&\underline{91.44}&90.15&\textbf{96.98}\\\hline
            	mj-rand&87.35&92.41&91.63&\underline{92.44}&\textbf{94.17}\\\hline
            	sdss1&88.96&88.35&87.74&\underline{90.36}&\textbf{91.24}\\\hline
            	sdss2&84.11&80.01&81.24&\underline{84.76}&\textbf{86.17}\\\hline
            \end{tabular}
         }
         \label{tab: prediction_justification}
        \end{center}
            \vspace{-1.3em}
    \end{table}
    
    In addition, simple models like LSTM and GRU, have restricted model capacity and limited number of learnable parameters, causing difficulties in effectively capturing complex relationships among features \cite{nnparameffect}. Thus, these models are not well-suited for a multi-variable problem with complex inter-relationships like semantic prefetching. \hblue{To tackle these defects, the LSTM should be used in a multi-layer structure such as encoder-decoder architecture.}

    The encoder-decoder LSTM (ED-LSTM) architecture has a superior ability to efficiently grasp the context and the dependencies between variables, making it one of the most popular architectures in natural language processing \cite{encdeclstm}. Its encoder maps the features into the latent states and inherently captures a hierarchical and high-level representation of the input sequence. Subsequently, the decoder, initialized with the encoder state, receives latent states values as the input and generates the output sequence. \hblue{We observe that compared to ML-LSTM, ED-LSTM, with its hierarchical representation and state initialization, can be more effective in capturing and leveraging the temporal dependencies in the input sequence.}

\begin{figure}
        \centerline{\includegraphics[width=0.88\linewidth]{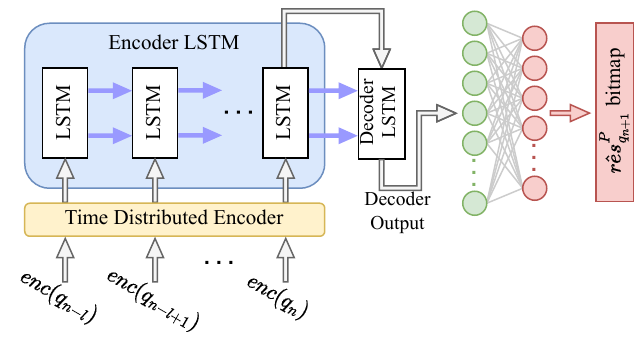}}
        \vspace{-1.2em}
        \caption{The prediction model encoder-decoder architecture}
        \label{fig_prediction_model}
         \vspace{-2.1em}
    \end{figure}

    Figure \ref{fig_prediction_model} depicts our prediction model with ED-LSTM architecture. \hblue{{\far} generates $enc(q)$ using the \emph{set} of partitions accessed by the queries, which remains invariant to the query plan and access order}. The model takes $\langle enc(q_{i})\rangle_{i=n-l}^n$ as input and compresses them using a time-distributed layer, resulting in a sequence of 128-dimensional vectors. These vectors are then fed into the ED-LSTM, designed with single-layer LSTMs. The decoder's output further passes through two dense layers, each containing $|P|$ units. The final dense layer generates $\hat{res}_{q_{n+1}}$, a $|P|$-dimensional bitmap vector, representing the probabilities of each partition getting accessed in the subsequent query. By leveraging these probabilities, we can select top $k$ partitions most likely to be requested next.

    With this model, we transform the prefetching problem into a multi-label classification task. Each neuron in the final dense layer corresponds to a specific partition (a.k.a. label), and a sigmoid activation function is applied to each neuron to ensure its output is a probability within the range of 0 to 1. Subsequently, we apply the binary cross-entropy loss function independently to each label and then aggregate them to calculate the total model loss. Considering $y_i \in \{0, 1\}$ and $\hat{y}_i$ as the true and predicted probability of partition $p_i$ being in the $res^P_{q_{n+1}}$, the loss function can be defined as follows:
    \begin{equation}
        \mathcal{L} = -\sum_{i} [y_i \cdot \log(\hat{y}_i) + (1 - y_i) \cdot \log(1 - \hat{y}_i)]\label{eq:loss_function}
      \vspace{-0.1em}
    \end{equation}
    
    To fine-tune the model after a repartitioning procedure, we start by freezing all layers up until the first dense layer. Subsequently, we proceed to fine-tune the system on the selected workloads for 15 epochs using a learning rate of $10^{-5}$.

\section{Evaluation}
    We evaluate {\far} against the state-of-the-art (SOTA) prefetchers across both SQL-based and navigational workloads gathered from SDSS DR7~\cite{abazajian2009sdss} and SQLShare \cite{sqlshare} real-world datasets. For completeness, we extend our evaluation to non-exploratory workloads by testing it on the TPC-DS benchmark. In this section, we first explain the experimental setup, then show the performance of {\far} against the SOTA prefetchers, and conclude the section with the sensitivity analysis of the parameters of interest.

 \subsection{Experimental Setup}
    
  \subsubsection{Implementation and Configurations}
    {\far} and the clay-based partitioning are implemented in python and the model is developed using TensorFlow/Keras framework \cite{tensorflow2015-whitepaper}. We configure each LSTM layer with 64 cells and conduct training using batches of size 32. Training continues until either early stopping based on the loss of validation data (10\% of the train data) or reaching the maximum number of training epochs, set at 75. For the prediction model, we employ Cross Entropy as the loss function, and for the autoencoder model, Mean Square Error is used. In both models, we utilize the Adam optimizer \cite{kingma2014adam} for the training process. 
    
    In our experimental setup, we incorporate {\far} as a middleware layer on top of the  PostgreSQL~\cite{stonebraker1986design} server. In practical terms, {\far} makes the prefetch decision and fetches the chosen blocks using the pg\_prewarm module within the PostgreSQL server.

    Unless mentioned otherwise, the configuration in the experiments are as follows: cache size = 4GB, partition size = 128 block, block size = 32kB, model input sequence length $l$ = 4, $k_w$ = 10, initial max partition load $\theta$ = 1, repartitioning threshold $l_p$ = 2500.

  \subsubsection{Datasets}   
  
    We use the seventh Data Release of Sloan Digital Sky Survey (SDSS DR7) \cite{abazajian2009sdss} dataset, which is an open-access database containing digital astronomy data, accessible via various tools including navigation and SQL search tools. Utilizing its publicly available workloads\footnote{http://cas.sdss.org/runs/en/tools/search/sql.asp}, we select two weeks of system logs (86 thousand entries) from WebLogs table for \textbf{navigational SDSS workloads}, and a subset of logs spanning three months (200 thousand queries) from SqlLogs table, for \textbf{SQL-based SDSS workloads}.
    
    The original DR7 database is hosted on SQL Server and has 20TB of data, comprising 95 tables and 51 views. For our experiments, we first migrate MyBestDR7\footnote{http://www.skyserver.org/myskyserver/}, a 2GB subset of DR7 to the PostgreSQL and extend its size by retrieving more data from the DR7 website, using SciScript\footnote{https://github.com/sciserver/SciScript-Python} python library. The experiments are performed on a 16GB dataset, comprising the same number of components.
    
    Our navigational SDSS database is a 4GB subset of our BestDR7 database, containing tables with attributes that form the data available in the SDSS navigation tool. The data is organized into four levels of granularity, each stored in a separate table.

    \hblue{We also test {\far} on two additional exploratory datasets gathered by the SQLShare \cite{sqlshare} project: \textbf{Birds}, containing 8GB of data and 6 tables including textual values in different languages, and \textbf{Genomics}, which contains 10GB of data spread across 13 tables.}

    Lastly, the \textbf{TPC-DS} database used in our evaluation is generated with a scale factor of 10.
    \hblue{We exclude the query templates that take more than ten minutes to complete and generate our test and train queries using the remaining 93 templates.}
    
  \subsubsection{Workloads}\label{sec:workloads}

    \paragraph{Exploratory SQL-based Workloads}\vspace{-0.1em} We classify the query workloads into three broad categories and within each category, considered two types: one with a recognizable pattern in the accessed data \textit{(regular)}, and the other without such patterns \textit{(random)}. The recognition of patterns is based on the sequence of accessed tables and the sequence of differences in the accessed block addresses (LBA-delta). In addition to these six categories, all of which consist of single-user sessions, our model is also tested against a full workload, encompassing queries from all categories executed by one or more users. The resulting types are as follows:
    \begin{itemize}
        \item \textbf{Single-user, single table access}: All queries in the \textit{single-regular (s-reg)} and \textit{single-random (s-rand)} workloads access just one particular table. Usually, workloads in \textit{s-reg} category are sequential table accesses.
    
        \item\textbf{Single-user, multiple table access without join}: Queries in the workload accesses a single table, while multiple tables are accessed in total across the entire workload. Workload is \textit{multiple-regular (m-reg)} if a recurrent pattern exist in the accessed LBA-delta, else \textit{multiple-random (m-rand)}.
    
        \item\textbf{Single-user, multiple table access with join}: Queries in this category of workloads mostly contain up to 4 join operations. If there is a pattern in the LBA-delta accessed by the queries, the workload is in type \textit{multiple-join-regular (mj-reg)}, and \textit{multiple-join-random (mj-rand)} otherwise.
        
        \item\textbf{Multi-user, full workload}: A \textit{full} session of a dataset (\textit{f-<dataset>}) with all types of queries, executed by one or more users
        . This mostly mimics real-life exploratory workloads.
    \end{itemize}
    
    \paragraph{Exploratory Navigational Workloads}\vspace{-0.1em} We use three types of exploration based on the gaps between consecutive viewed area~\cite{battle2016prefetching, doshi2003prefetching}.
     \begin{itemize}
        \item \textbf{Smooth exploration}: The user observes adjacent areas by panning to the neighboring region or zooming in and out. There is no significant gap between consecutive locations.
        \item \textbf{Jumping navigation}: This type involves \textit{Jumps} in the actions, meaning that the user requests an area that is not in the vicinity of the current viewport. In this type, the user observes some adjacent area and then jumps into another location to do the same.
        \item \textbf{Random navigation}: The user randomly probes data space \hblue{without adhering to specific patterns in spots coordinates.}
    \end{itemize}
    
\vspace{-0.3em}
    \paragraph{Non-exploratory SQL-based Workloads over TPC-DS} We generate a test workload, using 93 templates of this benchmark which 25\% templates are not used in generating training workloads. 

\vspace{-0.1em}
  \subsubsection{Baselines}\label{sec:baselines}
        We compare our proposed method on query workloads against traditional prefetching methods implemented in most mainstream DBMSs, and SOTA learning-based data prefetchers. 
    
    \begin{itemize}       
        \item \textbf{Lookahead} \cite{smith1978lookahead}: One block lookahead algorithm is the simplest prefetcher which is used in several DBMSs. When an address is accessed, it simply prefetches the next one. 
        
        \item \textbf{Random Readahead (Rand-Readahead)} \cite{opdenacker2007readahead}: Stores a window trace of recent demanded addresses. If within a window, a predefined number ($l_{RR}$) of blocks of an extent\footnote{https://dev.mysql.com/doc/refman/8.0/en/glossary.html\#glos} get accessed, the prefetcher retrieves the entire extent.
        
        \item \textbf{Naïve prefetcher}: It calculates the LBA-delta and selects its prefetching candidate blocks by recursively adding the most frequent LBA-delta to the last accessed address.
        
        \item \textbf{SGDP} \cite{2023sgdp}: This SOTA prefetcher represents interactive relations among LBA-delta streams using a weighted directed graph structure and learns the pattern of LBA-deltas using a gated graph neural network. 
    \end{itemize}

    In {\far}, we prefetch $k$ partitions with a maximum size of \textit{MaxParSize}. Thus, we extend Lookahead and Naïve models to prefetch $k\times$ \textit{MaxParSize} blocks instead of one. In the Rand-Readahead model, we configured extents to be as large as twice \textit{MaxParSize} and $l_{RR}$ to 13 which is its default value in MySQL Server. In SGDP, after each query, we recursively predict the next LBA-delta for $k\times$ \textit{MaxParSize} times to get the proper number of candidate partitions.
    
    On navigational workloads, we compare our method against a SOTA prefetcher, \textbf{ForeCache} \cite{battle2016prefetching}, which is a hybrid prefetcher working with both the history of the movements and the data semantic similarities of the adjacent area.
    
\vspace{-0.1em}
    \subsubsection{Metrics}
        As performance metrics, we employ hit ratio, presented in Equation (\ref{eq_hit_ratio}) and prefetch coverage presented in Equation (\ref{eq_pref_cover}). The prefetch coverage is determined by subtracting the number of cache misses in a system without a prefetcher (NP) from the number of cache misses in a system with a prefetcher. This metric quantifies the percentage of cache misses eliminated after implementing a prefetcher in the system (the higher the better). 
        \vspace{-0.2em}
        \begin{equation}
            \begin{split}
                Hit\ Ratio = \frac{Hits}{Hits +  Misses}\label{eq_hit_ratio}
            \end{split}
        \end{equation}
        \begin{equation}
            Coverage = \\ \frac{Misses_{NP} - Misses}{Misses_{NP}}\label{eq_pref_cover}
        \end{equation}

        Additionally, we measure the I/O time to illustrate the impact of each model on reducing query response time. Query execution time can be divided into I/O time and computation time, with prefetching techniques only affecting the former and having no impact on the computational overhead of the queries. Hence, prefetchers can improve the query response time to a limit, which is when it gets 100\% hit ratio. \hblue{ To better demonstrate the I/O time improvement, we report the relative I/O time calculated as follows:}
        \begin{equation}
            \begin{split}
                Relative\ t\_io_{pr} = \frac{t\_io_{pr}}{t\_io_{NP}}\label{eq:relative_exec_time}
            \end{split}
        \end{equation}
        where $t\_io_{pr}$ is the I/O time of the workload when employing $pr$ prefetcher, and $t\_io_{NP}$ stands for the I/O time of the NP system.
        
\vspace{-0.2em}
    \subsubsection{Hardware}
        All experiments are conducted on a server running Ubuntu 18, equipped with 48 Core at 2.4GHz, 1.1TB RAM, 50TB disk (10K RPM), and one NVIDIA V100 GPU with 16GB memory.
  
  \vspace{-0.1em}  
 \subsection{Results}
 \subsubsection{{\far} versus the SOTA} We test {\far} on different workloads and compare its performance against the baselines.
 
\begin{figure*}
      \centerline{\includegraphics[width=0.90\textwidth]{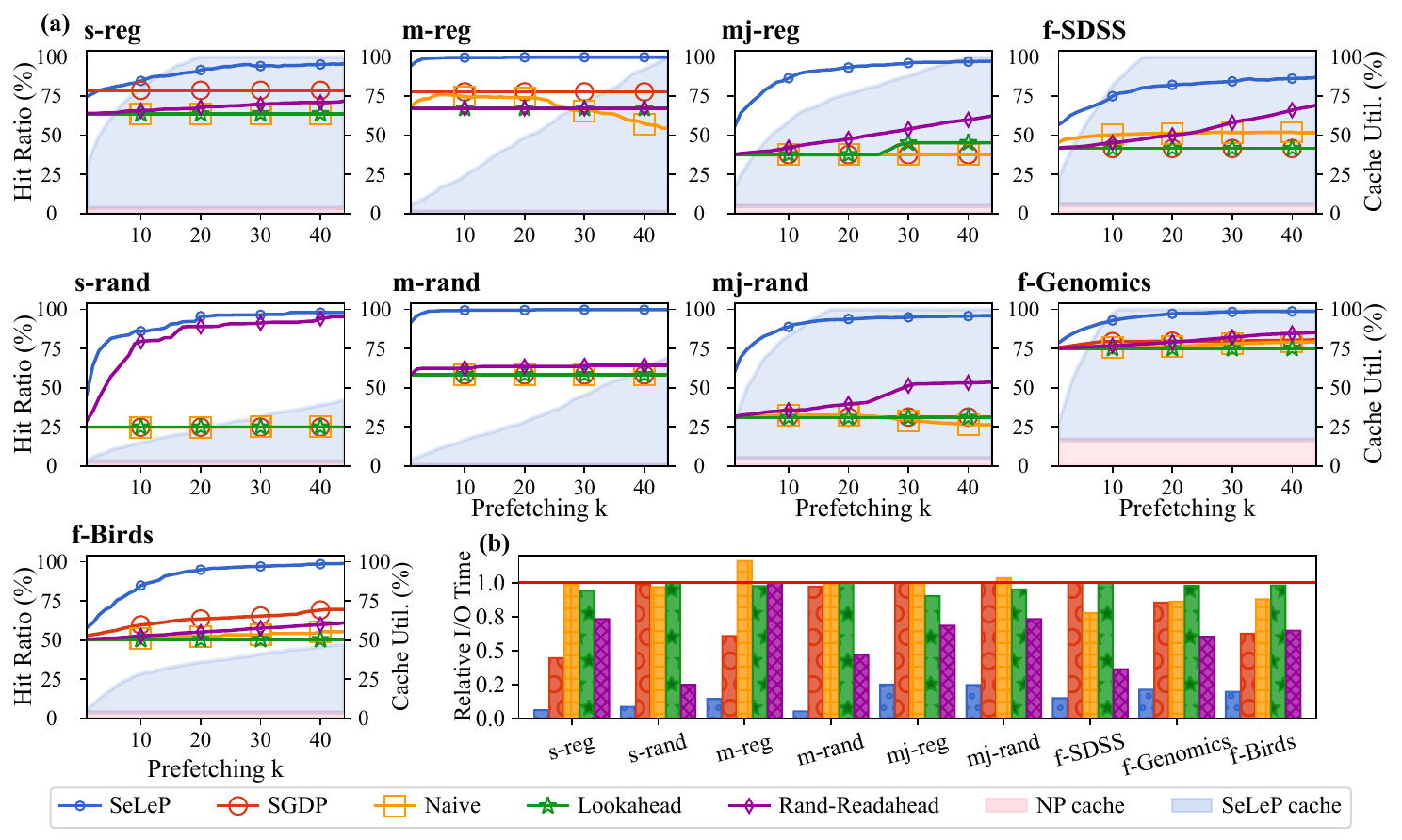}}
        \vspace{-0.9em} 
        \caption{\hblue{(a) Hit ratio in SQL-based exploratory workloads. The shaded area shows the portion of cache utilized by the workload (pink) and the additional portion consumed by {\far} prefetcher (light blue). (b) Relative I/O time with $k$=42 for same workloads.}}
        \label{query_hitrate}
        \vspace{-0.7em} 
    \end{figure*}

    \paragraph{\textbf{SQL-based workloads}}
    In this subsection, we report the results of the tests conducted on nine SQL-based workloads (\S \ref{sec:workloads}) to evaluate both {\far} and the baselines (\S \ref{sec:baselines}). Figure \ref{query_hitrate}(a) shows the hit ratio of {\far} and other baselines at various $k$, informing our choice of $k=42$. In \autoref{tb:sql_coverage} and \autoref{query_hitrate}(b), we present the miss coverage and I/O time results for this selected $k$, respectively.
    
    In the simplest workload, \textit{s-reg}, SGDP initially achieves the highest hit ratio near 80\%. However, on more complex yet regular workloads, {\far} outperformed SGDP by a significant margin. SGDP could perform better in smaller $k$ due to its individual block selection, where subsequent requested blocks, even in different partitions, can be found among the first few \textit{MaxParSize} prefetched blocks. However, for larger $k$ values, {\far} proves more effective in bringing more accurate data to the cache and SGDP is incapable of predicting diverse block addresses and has a constant hit ratio. 

    SGDP halts prefetching upon detecting uncertainty in the output predictions, which typically occurs when it encounters irregular LBA-deltas. Irregular LBA-deltas refer to values other than the top 1000 frequently occurring LBA-delta in the training workload. The presence of irregular LBA-delta is a common phenomenon in data exploration with large datasets and random data access. Furthermore, SGDP frequently predicts zero as the next LBA-delta multiple times in a row. This leads to numerous duplicates LBAs among the selected prefetch candidates, which is the primary reason why SGDP achieves near zero miss coverage in multiple tests.

    Conversely, Naïve prefetcher, which has a similar logic to select prefetch candidates, continues to generate and prefetch block addresses without considering memory consumption or any uncertainty. Consequently, in larger $k$, its hit ratio decreases and it attains negative miss coverage in \textit{m-reg} and \textit{mj-rand} tests. This occurs since the previously accessed blocks that are still needed are evicted from the cache to make space for the prefetched blocks.

    \begin{table}[htbp]
        \caption{Miss Coverage in SQL-based Workload with $k = 42$.}
        \vspace{-0.9em}
        \begin{center}
            \scalebox{0.9}
            {   
            \begin{tabular}{cccccc}
                Test & Naïve & Lookahead & \small{Rand-Readahead} & SGDP & SeLeP \\
                \hline
                s-reg & 0.0 & 1.25 & 31.97 & 49.31 & \textbf{79.28} \\
                s-rand & 1.01 & 0.0 & 93.67 & 0.71 & \textbf{97.26} \\
                m-reg & -33.62 & 0.54 & 0.14 & 32.16 & \textbf{99.42} \\
                m-rand & 0.0 & 0.0 & 14.41 & 1.28 & \textbf{99.42} \\
                mj-reg & 0.0 & 12.17 & 37.58 & 0.0 & \textbf{95.51} \\
                mj-rand & -6.54 & 2.41 & 32.38 & 0.53 & \textbf{91.8} \\
                f-SDSS & 17.17 & 0.0 & 44.61 & 0.0 & \textbf{69.22} \\
                f-Genomics & 16.88 & 0.86 & 39.84 & 16.93 & \textbf{94.82} \\
                f-Birds & 1.01 & 0.37 & 20.33 & 40.0 & \textbf{97.61} \\
                \hline
            \end{tabular}    
            }
        \label{tb:sql_coverage}
        \end{center}
        \vspace{-1.5em}        
    \end{table}

    In workloads with \textit{random} access pattern, which is more common in data exploration, {\far} always attains the best hit ratio, while SGDP and most of the traditional prefetchers could not bring any particular advantages to the system. Rand-Readahead has a high hit ratio in \textit{s-rand} since the data is randomly accessed from a single table, making it an ideal scenario for this model which does not rely on patterns to trigger prefetching. Nevertheless, in \textit{m-rand} and \textit{mj-rand}, where each query accesses data randomly from multiple tables, there are not enough common extent blocks in the window trace so the model triggers the prefetching process.

    It is evident that {\far} can benefit different types of exploratory workloads, in particular the ones consisting of random access patterns or queries with joins. Workloads falling under the \textit{s-reg} and \textit{m-reg} categories mainly consist of sequential data accesses. The lower hit rate of {\far} in sequential accesses at small $k$ values is primarily due to its shuffling of blocks in partitions based on workloads. Therefore, if certain types of workloads, such as sequential accesses, are less frequent compared to others, it is likely that their order is not preserved effectively in the partitions.

    \begin{table*}
        \begin{minipage}[b]{0.74\linewidth}
        \centerline{\includegraphics[width=0.98\linewidth]{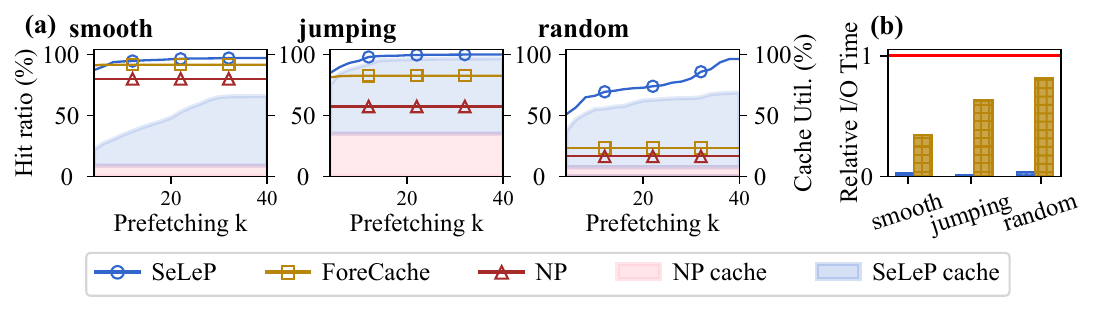}}
            \vspace{-1.3em}
        \captionof{figure}{(1) Hit ratio in navigational workloads. The pink area represents portion of the 500MB cache, utilized by the workload, and the light blue area is the additional portion consumed by {\far}, (2) Relative I/O time of {\far} and ForeCache for the same workloads with $k=36$.}
                \label{fig:navi-hitrate-exec}
                \vspace{-1em}  
        \end{minipage}\hfill
        \begin{minipage}[b]{0.25\linewidth}
        \centering
            \caption{Miss Coverage of {\far} and ForeCache in Navigational Workloads with $k = 36$.}
            \scalebox{0.99}{
            \begin{tabular}{ccc}
                Test & ForeCache & SeLeP\\
                \hline
                smooth & 58.21 & \textbf{85.88}\\
                jumping & 59.26 & \textbf{99.93}\\
                random  & 8.16 &\textbf{85.44}\\
                \hline
            \end{tabular}}
            \label{navi_coverage}
            \vspace{2.2em}
        \end{minipage}
    \end{table*}

    Figure \ref{query_hitrate} also provides insights into the portion of the cache populated by the workload (pink) and the additional portion consumed by {\far} (blue). From this data, we conclude that even when the cache is full, e.g. in \textit{f-SDSS} and \textit{f-Genomics}, the prefetcher can effectively populate the cache and enhance the hit ratio.

    \autoref{tb:sql_coverage} signifies that in all workloads, {\far} has the highest coverage, meaning that it is more capable than the other models to eliminate cache misses from an NP system. In cases where coverage is zero, it means that the prefetcher could not accurately predict any upcoming access that is not already in the cache and therefore its performance is the same as an NP system. For instance, Lookahead hit ratio is equal to the NP hit ratio and it has almost zero miss coverage in all workloads except the \textit{mj-reg} test.

    Figure \ref{query_hitrate}(b) depicts the relative I/O time, calculated using Equation (\ref{eq:relative_exec_time}). In this plot, 1 implies the maximum I/O time, equivalent to NP system that solely utilizes the LRU cache to store previously accessed blocks. Conversely, reaching 0 means the lowest possible I/O time, similar to a model that attains a 100\% hit rate on every query of the workload. If the relative I/O time of a system is 0.2, \hblue{it conveys that the system was able to reduce the I/O time by 80\%.} Thus, in this plot, a lower value signifies better performance.

    While the hit ratios in workloads such as \textit{f-Genomics} workload, are close, there is a significant variance in the I/O times of each prefetcher. Across all tests, {\far} has the lowest relative I/O time, yielding a reduction of up to 95\% in I/O time for single-table workloads and approximately 70\% in more complex multi-join workloads. The Naïve's negative coverage results in a relative I/O time exceeding 1, indicating it would actually increase the I/O time of the workload. Lookahead has a negligible impact on I/O time, same as  SGDP in most workloads except \textit{s-reg} and \textit{m-reg}. Among the baselines, Rand-Readahead performs the best, but it still falls short of achieving more than a 75\% reduction.

    It is worth noting that the access time of a group of blocks does not only depend on the quantity of the blocks; it is also influenced by the specific locations of these blocks on the disk. For instance, accessing ten blocks stored sequentially on the disk is significantly faster than retrieving ten blocks from various tables located in different parts of the disk. Conversely, the hit ratio is calculated based on the count of the presence or absence of the accessed blocks in the cache. Hence, achieving a hit ratio does not linearly translate into the same reduction in I/O time. It is also possible that systems with the same hit ratio have different I/O times, as the specific blocks that are missing and their location can significantly impact performance. This effect becomes particularly evident in complex workloads, such as \textit{full} and \textit{multi-table-joins}, which demand accessing blocks from various tables scattered across the disk. 

    Note that prefetching occurs during idle periods between user requests, terminating upon receiving a new query to avoid delaying execution. However, in the SDSS workload, the average prediction and prefetching process of {\far} is 64(ms)  which is far less than the average delay between user actions, ensuring completion before new requests arrive. All time overheads are reported in Table \ref{tab:times}.
  \vspace{-0.2em}  
\paragraph{\textbf{Navigational Workloads}}
    We compare {\far} and ForeCache using the mentioned performance metrics on navigational workloads. We treat each navigational query as a simple SQL query selecting data points within the viewport's boundary, and use these converted queries to train and test {\far}. Due to the smaller size of this database, we adjust some parameter values, setting \textit{MaxParSize} to 64, the block size to 16kB, and the cache size to 500MB.


     Figure \ref{fig:navi-hitrate-exec}(a) shows the hit ratio of the models on the navigational workloads. ForeCache performs well in smooth navigation where users mostly move to adjacent areas. However, in random navigation where users select different areas with random coordinates, it struggles to predict accesses accurately, leading to a low hit ratio. Since ForeCache only examines the adjacent area of the current viewport, increasing $k$ does not notably improve its performance.
    
    Conversely, {\far} \hblue{detects the potential semantic relationship among the observed areas even if they have randomly changing coordinates.} It achieves a high hit ratio in all workloads and outperforms ForeCache in almost all cases, except for when $k$ is small in smooth navigation, which can be attributed to the same reason it does not perform optimally in sequential SQL-based workloads.

    Using Figure \ref{fig:navi-hitrate-exec}(a) we select $k=36$ to conduct other tests. Table \ref{navi_coverage} represents the coverage of {\far} and ForeCache in the navigational workloads with $k=36$, in which is evident that {\far} can more effectively eliminate cache misses from an NP system.
    
    Figure \ref{fig:navi-hitrate-exec}(b) depicts relative I/O time for {\far} with $k$ = 36 and ForeCache. While ForeCache reduces I/O time by about 60\% in smooth navigation and less than 40\% in random and jumping scenarios, {\far} achieves over 90\% reduction in all workloads. The converted navigational queries are structurally simple with no joins, often accessing adjacent blocks, leading to faster disk read for missing blocks compared to the SQL-based exploratory workloads. This clarifies Forecache's 20\% reduction in I/O time with a 25\% hit ratio.

    
  \paragraph{\textbf{Non-exploratory Workloads}}
    Though {\far} is designed based on characteristics of exploratory workloads, we conduct tests on non-exploratory workloads to underscore its effectiveness in such scenarios. We specifically choose TPC-DS benchmark, which features the largest schema and the most complex queries among other benchmarks, such as TPC-H. Given that these benchmarks share similarities in data and certain workload characteristics, we expect similar performance for {\far} across the other benchmarks.

    The key distinction between TPC-DS and exploratory workloads is the number of blocks accessed by each query. On average, each query in TPC-DS accesses about 40,000 blocks, while this number for SDSS is about 47. This is why $k$ has larger values on this test. Also, accessing a large number of blocks increases the likelihood of block reaccess, resulting in a high hit ratio for the NP system. 

\begin{figure}[htbp]
        \centerline{\includegraphics[width=0.95\linewidth]{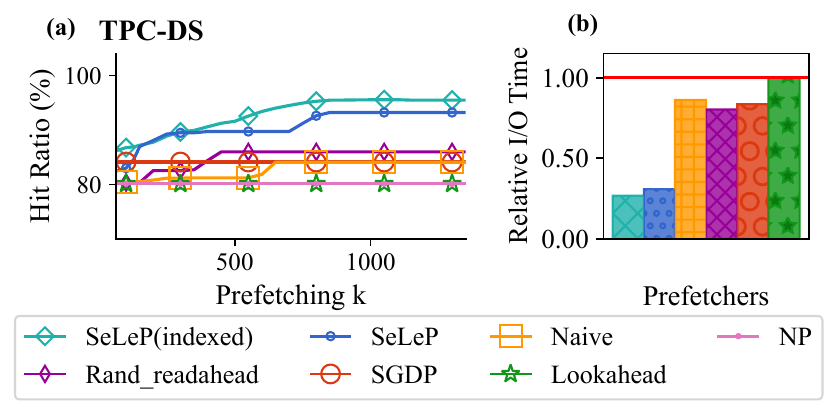}}
         \vspace{-1em}  
        \caption{\hblue{Hit ratio and relative I/O time for TPC-DS non-exploratory workload with 40\% novel templates and $k = 900$. The workload cache utilization is 100\% and is not shown.}}
        \label{fig:tpcds_hitrate_exec}
           \vspace{-1.2em}
    \end{figure}

    \hblue{Figure \ref{fig:tpcds_hitrate_exec}(a) depicts hit ratio of {\far} and other baselines and Figure \ref{fig:tpcds_hitrate_exec}(b) represents their Relative I/O time for $k$=900. Similar to the exploratory workloads, Lookahead prefetcher fails to cover any misses and essentially has the hit ratio and I/O time equivalent to the NP system. The performance of other baselines are similar, improving the hit rate and I/O time up to 6\% and 20\% respectively.}
    
    In contrast, {\far} can more effectively prefetch the required blocks and reduce the I/O time. It consistently achieves the highest hit ratio across nearly all values of $k$, except in certain cases where SGDP attains a higher hit ratio. However, as $k$ grows larger, SGDP maintains a constant hit rate, while {\far} achieves a 93\% hit ratio and 70\% I/O time reduction. Unlike Rand-Readahead and other approaches, {\far} enhances I/O time efficiency by prefetching data from various tables located across diverse disk sections, rather than just focusing on blocks local to the recently accessed ones.
    
    \paragraph{\textbf{Impact of physical design}} We explore the impact of physical database design by indexing the TPC-DS database and retesting with {\far}. For index tuning, we utilize HMAB \cite{hmab}, a SOTA database tuning tool, which generates new indices adaptive to the workload. Performance of this new prefetcher, named {\far}-indexed, is depicted in Figure \ref{fig:tpcds_hitrate_exec}. With the help of indices, the number of blocks getting accessed is reduced, which increases the chance of accurate access prediction, causing {\far}-index to have a higher hit ratio. For the same result, the hit ratio of the NP system on the enhanced database is 3.5\% higher than the NP hit rate on the original database. Additionally, in Figure \ref{fig:tpcds_hitrate_exec}(b), we have calculated the {\far}-index I/O time relative to the I/O time of the NP system in the same database. The results in Figure \ref{fig:tpcds_hitrate_exec} demonstrate that {\far} can further enhance performance even when there are changes in the database indices.

\subsubsection{Adaptivity test}\label{sec:adaptivity_test}
    As mentioned earlier, exploratory workloads can alter due to the users' evolving knowledge or changing interests and an effective exploratory tool must be able to adapt to these shifts. While it is uncommon for a large portion of the exploratory workload to undergo abrupt shifts very frequently, we conduct an adaptivity test to demonstrate {\far} can adjust to such workload shifts. \hblue{In this test, after every $l_p=500$ queries, system repartitions, taking 102.3(s) on average, and the prediction model fine-tunes with the recent queries which, due to small $l_p$, takes 3.61(s) on average.}
    


    The initial batch of queries, starting from sequence number (SN) 0, constitutes a simple session without unseen blocks (blocks never accessed before), or new templates. In the next batch, SN=2000, 80\% of the accessed tables change, accompanied by 20\% unseen blocks. At SN=4000, 75\% of the requested tables change, and 33\% unseen blocks are accessed using 66\% new templates. The most challenging batch starts at SN=6000, where all accessed tables change, and 85\% unseen blocks are accessed using completely new query templates.

\begin{figure}[htbp]
        \centerline{\includegraphics[width=0.99\linewidth]{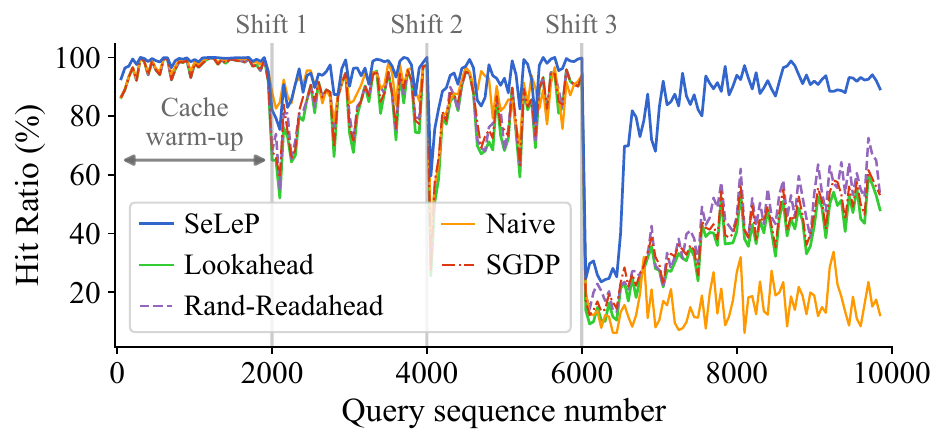}}
        \vspace{-1em}
        \caption{Hit ratio of prefetchers for each 50 consecutive queries in a shifting workload.}
        \label{fig:adaptivity_test}
        \vspace{-1.2em}
    \end{figure}

    Figure \ref{fig:adaptivity_test} depicts the hit ratio of the prefetchers for each 50 consecutive queries. It is evident that {\far} swiftly adapts to workload shifts, and maintains a more consistent hit rate compared to other approaches. Upon receiving completely new workloads in the third shift, its hit rate drops less steeply than other approaches, and after receiving $l_p$=500 queries and completing repartitiong and fine-tuning, its hit rate improves by 50\%. Same as other tests, Lookahead and SGDP exhibit nearly identical behaviour to the NP system. Random and Naïve models occasionally achieve better hit ratios in certain batches. Following the last shift, the Naïve prefetcher however overfills the cache with incorrect prefetch decisions achieving the lowest hit ratio. Due to stochastic nature of the last query batch, neither of the prefetchers could converge to a steady hit ratio.

    \begin{figure*} [htb]
        \includegraphics[width=0.98\linewidth]{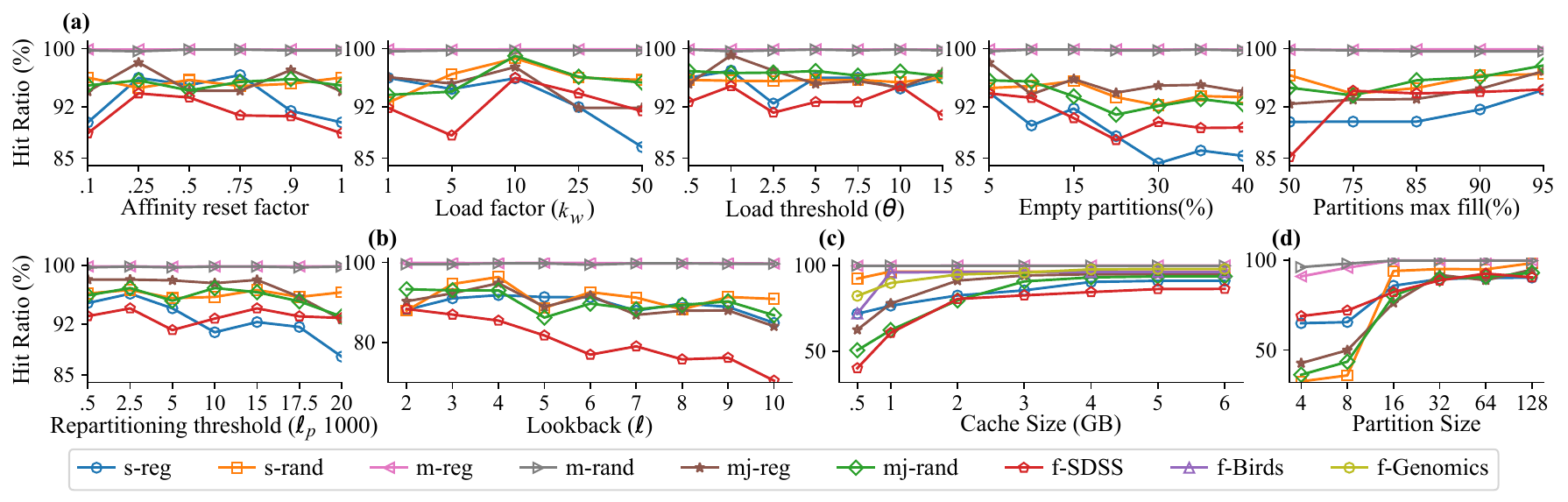}
    \vspace{-0.8em}
    \caption{Effect of (a) partitioning config, (b) lookback $l$, (c) cache size, and (d) $MaxParSize$ on hit ratio of SQL-based workloads.}
    \label{fig:sensitivity}
    \vspace{-0.8em}
\end{figure*}
    

   

  \subsubsection{Sensitivity analysis}\label{sec:sensitivity_test}
  {\far} has several configuration parameters mostly related to its partitioning algorithm. To address the challenges associated with these parameters, we run clay-based partitioning and the prediction model with varying configurations, evaluating their impact on the hit ratio of the SQL-based workloads.

    \paragraph{Partitioning parameters} The initial value for $\theta$ should be a small number to enforce stricter repartitioning in the beginning that the partitions are a group of sequential blocks. As the repartitioning procedure progresses, this value gradually increases. As explained in \S \ref{sec:data_partitioning}, we have restricted the partitioning algorithm to not creating new partitions, and instead increase the $\theta$ value in case adding a new partition is the only option. A large $l_p$ results in multiple partitions having high loads and sticking in the situation that $\theta$ should be modified. Such increases in the $\theta$ value reduce the effectiveness of the partitioning and lead to irrelevant blocks grouping together.
 
    Figure \ref{fig:sensitivity}(a) illustrates how the hit ratio changes as each of clay-partitioning  parameters is modified. Based on these results, we selected the following parameter values: $l_p=2500$, $k_w=10$, and $\theta = 1$. In the initial partition generation phase, each partition should be filled to 95\% of its capacity, and an extra $0.05 \times |P|$ empty partitions should be added to the system. After each repartitioning, the weights of the affinity graph should be reduced by 0.25.
    
 \vspace{-0.2em}
    \paragraph{Lookback} The line charts in Figure \ref{fig:sensitivity}(b) display the hit ratio of {\far} as the prediction model is trained with varying lengths of query encoding sequences for input. This parameter serves as a history size, and from the results, we can infer that larger history sizes can have a negative impact on the system's performance. The optimal choice for this parameter is found to be 3 or 4 and we decided to set this variable to 4.
 \vspace{-0.2em}  
    \paragraph{Cache size} To show effectiveness of {\far} with different cache sizes, we have conducted a test with all exploratory workloads with $k=24$ and cache sizes varying from 500MB to 6GB. Results in \ref{fig:sensitivity}(c) illustrate that  hit ratio is lower for smaller cache sizes while the system could make improvement with a 1GB or larger cache. However, based on the workload data footprint,  the system can achieve a high hit ratio even with 500MB cache. For instance, in \textit{s-rand}, \textit{m-reg}, \textit{m-rand} and \textit{f-Birds} tests with 500MB cache, the system yields 40\% improvement in  hit ratio on average over the NP.

\vspace{-0.2em}
    \paragraph{Maximum partition size}\label{maxparsize}
    To provide a more comprehensive representation of the impact of partitioning, we conducted tests with {\far} using variable partition sizes. We adjust the value of $k$ based on the partition size to prefetch an equivalent number of blocks in all systems. Figure \ref{fig:sensitivity}(d) illustrates the hit ratio relative to the maximum partition size. With larger partitions, usually, there is a greater likelihood of accurately predicting subsequent accesses, even when prefetching the same number of blocks. This could also be observed in Figure \ref{query_hitrate} where heuristic methods such as Naïve and Rand-Readahead prefetch the same number of blocks but achieve significantly lower hit ratios compared to {\far}. 
    \begin{table}[htbp]
    \caption{Average time of block encoding, data preparation (including block encoding), access prediction, prefetching selected blocks, and repartitioning in different datasets.}
    \vspace{-0.5em}
    \begin{center}
        \scalebox{0.78}{
        \begin{tabular}{|c|c|c|c|c|c|}
            \hline
             Action & Navigational & SDSS & Birds & Genomics & TPC-DS \\
            \hline
            \centered{Block encoding\\(s) one-off }& 779.6 & 2898.17 & 3672.62 & 1997.86 & 4978.88 \\
            \hline
            \centered{Data preparation\\(s) one-off } & 1108.11 & 3848.83 & 8548.37 & 3220.22 & 7358.85 \\
            \hline
            \centered{Prediction (ms)} & 2.16 & 8.28 & 11.08 & 18.44 & 356.15 \\
            \hline
            \centered{Prefetching (ms)} & 20.87 & 55.81 & 44.94 & 52.05 & 511.52 \\
            \hline
            Repartitioning (s)& 17.31 & 142.81 & 228.26 & 456.92 & 1901.83 \\
            \hline
        \end{tabular}
     }
     \label{tab:times}
    \end{center}
\vspace{-1.5em}
\end{table}It is worth noting that despite prefetching a larger amount of data with larger partition sizes, the prefetch time remains negligible.

\subsubsection{Time overheads}\label{sec:costs}

    Table \ref{tab:times} represents the time overhead of different components of the system for each dataset. The data preparation as a one-off task typically requires the longest time, while partition access prediction and data prefetching are done at the level of milliseconds. Therefore, the prefetching process can lower the I/O time without delaying execution of the user's next query.

\section{Related Work}
    \paragraph{\textbf{Visual tools prefetchers}} are utilized to prefetch spatial data observed through a fixed-size viewport, such as a laptop screen \cite{battle2016prefetching, tauheed2012scout, atlas, kalinin2014interactive, doshi2003prefetching, tauheed2012scout, wan2018learning}. Their primary assumption is that users navigate smoothly through the data space, and select prefetch candidates from the neighboring area of the recently accessed region. These prefetchers make decision by repeating user's previous movements (\textit{action-based})\cite{atlas, doshi2003prefetching}, or learning and predicting  movement or location sequences patterns (\textit{learning-based})\cite{wan2018learning, tauheed2012scout}, or, akin to semantic prefetchers, exploit the similarity in data properties among data tiles to make predictions \textit{(characteristic-based)} \cite{battle2016prefetching, wan2018learning}. 

    \paragraph{\textbf{SQL-based tools prefetchers}} are mostly traditional prefetchers such as Random Readahead \cite{opdenacker2007readahead} and One Block Lookahead \cite{smith1978lookahead} which struggle with complex data access patterns. The only learning-based database prefetcher \cite{chen2021revisiting} applies offline training on a single set of data and falls short in meeting the adaptivity demands of exploratory tools. Memory prefetchers, extracting LBAs \cite{ki2000stride} and LBA-deltas~\cite{2023sgdp, 2020delta_lstm} sequece patterns, may find utility in SQL-based tools. SGDP \cite{2023sgdp} is the SOTA learning based memory prefetcher, demonstrating superior performance compared to other models. These prefetchers excel in memory systems where applications access data within their limited allocated memory portion and the LBA-deltas are less diverse, but struggle in data exploration due to the complexity and diversity of data access patterns.  

\vspace{-0.1em}
    \paragraph{\textbf{Workload-driven approaches for latency reduction}} Physical design tuning is a well-studied area where tuning decisions are made to optimize system performance based on resource constraints and data access patterns. These decisions involve adaptively creating or altering indices to reduce the query execution time \cite{hmab, dbabandit, tkde23, learnedsigmod23, icde24} or selecting which views to materialize or which results to cache to prevent repeated computation in the future \cite{hmab, OM1, OM2}. We view such approaches as complementary, as they mainly optimize computation part of execution (rather than I/O).
    \vspace{-0.3em} 
    
    \paragraph{\textbf{Data partitioning}} Partitioning is a technique to improve manageability of the data or load balancing in distributed systems~\cite{vo2014sato, wu2011partitioning}. Combined with other techniques such as index tuning for individual partitions~\cite{olma2017slalom} or locating partitions of frequently co-accessed data sequentially on the disk~\cite{pavlovic2016odyssey}, this approach can also be effective in reducing system latency and improving performance.

 \vspace{-0.3em}
\section{Conclusion}
    In this work, we present {\far}, a learning based semantic prefetcher to enhance interactivity of the  exploratory workloads. Unlike the majority of SOTA prefetchers, {\far} utilizes data semantics acquired through encoding data blocks rather than relying on logical data addresses. Our experimental results on several SQL-based, navigational and non-exploratory workloads indicate that prefetching based on data semantics can yield comparable or even superior performance (up to 40\% higher hit ratio in challenging workloads) when compared to prefetchers that rely solely on data addresses. Additionally, decision making based on the encoded query results calculated using data block encodings allows our prefetcher to cater to both types of workloads effectively. Furthermore, our system outperforms traditional address-based prefetchers in terms of speed (up to a 45\% greater reduction in I/O time), enabling faster and more responsive query execution, crucial for exploratory systems.
 \vspace{-0.3em}
\begin{acks}
We gratefully acknowledge support from the Australian Research Council Discovery Early Career Researcher Award DE230100366, and L’Oreal-UNESCO For Women in Science'23 Fellowship.
\end{acks}



\balance
\bibliographystyle{ACM-Reference-Format}
\bibliography{refs}

\end{document}